\documentclass[a4paper,20pt]{article}
    \usepackage[T1]{fontenc}
    \usepackage{graphicx}
    \usepackage{epsfig}
    \usepackage{amsmath}
    \usepackage{amsfonts}
    \renewcommand{\abstract}{}
\begin{document}
\makeatletter
\renewcommand{\@oddhead}{\textit{YSC'14 of Contributed Papers} \hfil \textit{A.V. Karnaushenko, E. Yu. Bannikova, V. M. Kontorovich}}
\renewcommand{\@evenfoot}{\hfil \thepage \hfil}
\renewcommand{\@oddfoot}{\hfil \thepage \hfil}
\fontsize{11}{11} \selectfont

\title{Frequency Dependence of Radio Images of Supernova Remnants}
\author{\textsl{A.V. Karnaushenko$^{1}$, E. Yu. Bannikova$^{1,2}$, V. M. Kontorovich$^{1,2}$}}
\date{}
\maketitle
\begin{center} {\small $^{1}$ Institute of Radio Astonomy NAS of Ukraine \\
$^{2}$ Karazin Kharkiv National University\\
a\_karnaushenko@mail.ru, bannikova@astron.kharkov.ua,
vkont@ira.kharkov.ua}
\end{center}

\begin{abstract}
Radio images of supernova remnants  in the framework of diffusion
model are discussed. The distribution profiles  of synchrotron
radiation intensity for spherical injection source of relativistic
electrons are reduced at different frequencies. An explanation of
the observational data obtained on UTR-2, according to which the
size of the supernova remnant at decametric waves is larger than the
remnant size  at high frequencies, is given.
\end{abstract}

\section*{Introduction}
\indent \indent The radio images of the supernova remnants (SNR)
obtained with the UTR-2 radio telescope by M.A. Sidorchuk and E.A.
Abramenkov show the difference between  SNR sizes at low frequencies
(decametric waves) vs. the higher ones \cite{1}.

For example, the brightness temperature contour map of SNR HB3 at
the frequency 20 MHz is shown in Fig.1. The temperatures on the
isophotes are given in $10^3 K$. At the lower corner the UTR-2
directional diagram is shown as a dashed circle. The center of SNR
is denoted by the white $"+"$. At the frequency 1420 MHz radio image
of SNR corresponds to the size  $60\times80$ pc (the region limited
by a white circumference)\cite{2}. At frequency 20 MHz it
corresponds to the size  $70\times95$ pc \cite{1}. We can see that
the SNR size at low frequency is larger than the SNR size at high
frequency. In this work we propose an explanation of the SNR size
changing with frequency in the framework of diffusion model.

\section*{Diffusion model}
\indent \indent In the diffusion model we consider the SNR as a
spherical source of relativistic electrons. The particles are
accelerated by the spherical shock wave (SW) front and next further
propagate by means of diffusion into the environment medium  losing
their energy due to synchrotron radiation \cite{3}. The basic
equation is the kinetic equation (KE) for electron distribution
function (EDF) ${N(E,t,\vec{r})}$ with a source $Q(E,t,\vec{r})$ and
with diffusion as a mode of the electron propagation

\begin{equation}\label{eq1}
  \frac{\partial N}{\partial t} +
                         \frac{\partial(B(E)N)}{\partial E} -
                         D\Delta N
                        = Q(E,t,\vec{r}),
\end{equation}
were $\Delta$ is the Laplace operator, ${D}$ is the diffusion
factor that as we will show bellow can be chosen to have no energy
dependence.

The second term in KE describes the synchrotron losses of electron
$$B(E)=-\beta E^2, \quad \beta=\frac{32
  \pi}{9}\left(\frac{e^2}{mc^2}\right)^{2}\cdot\frac{W_H}{m^2c^3},$$
where $e$ is the electron charge, $m$ is the electron mass, $c$ is
the light velocity and $W_H=H^2/(8 \pi)$ denotes the energy density
of the magnetic field.

The right part of equation (\ref{eq1}) corresponds to the source
of relativistic electrons: $${Q(E,t,\vec{r})=Q_{0}\cdot
N_{inj}(E)\cdot S(\vec{r})\cdot\Theta(t)},$$ where
$S(\vec{r})=\delta(|\vec{r}|-R)$  sets the spherical form of the
source, with radius $R$,
$N_{inj}=E^{-\gamma_0}\cdot\Theta(E_{max}-E)\cdot\Theta(E-E_{min})$
is the distribution function of injected electrons,
$\gamma_{0}=2$, $\Theta(...)$ is the Havyside's step function.

The solution of KE is the EDF for the spherical source \cite{3},
\cite{4}:
\begin{equation}\label{eq2}
  N(E,t,\vec{r})=\frac{Q_0}{(4 \pi)^{3/2}\cdot E^2}\cdot\int_{-\tau_+ (
  E,t)}^{\tau_-(E,t)}d\tau_{-} '\cdot\frac{\tilde{E^2}(\tau_{+},\acute{\tau_-})}{\alpha^{3/2}(\tau_{+},\acute{\tau_-})}\cdot N_{inj}(\tilde{E})
  \cdot R^2\int\ d \Omega\cdot  exp\left(\frac{(\vec{r}-\vec{r}')^2}{4\cdot\alpha(\tau_{+},\acute{\tau_{-}})}\right),
\end{equation}where
$$\tau_{\pm}(E,t)=\frac{1}{2}\cdot\left(t\pm\frac{1}{\beta}\left(\frac{1}{E_{max}}-\frac{1}{E}\right)\right),
\quad
\alpha(\tau_{+},\acute{\tau_{-}})=D_{0}\cdot(\tau_{+}+\acute{\tau_{-}}),$$

$$\tilde{E}(\tau_{+},\acute{\tau_{-}})=\frac{1}{\frac{1}{E_{max}}-\beta\cdot(\tau_{+}-\acute{\tau_{-}})},$$
$(\vec{r}-\vec{r}')^2=(x-x')^2
  +(y-y')^2+(z-z')^2,\quad
  x'=R\cdot\sin\theta\cdot\cos\varphi
  ,\quad y'=R\cdot\sin\theta\sin\varphi ,\quad
  z'=R\cdot\cos\theta,\quad d\Omega=\sin\theta  d\theta  d\varphi$.

The physical parameters of the diffusion model are the following: he
life time ${\tau=1/(\beta\cdot E)}$ - the time of particle existence
with energy $E$, diffusion length ${l_{dif}=\sqrt{D_{0}\cdot\tau}}$
- the distance, which  particle passes during the life time, and the
diffusion velocity is $V_{dif}=D_{0}/l_{dif}$. The diffusion length
defines the size of the radio image of SNR. Knowing the EDF, we can
calculate the intensity distribution of the synchrotron radiation
$I$ \cite{5}. The intensity  distribution in the image plane for the
sources resolved by the radio telescopes is given by:

\begin{equation}\label{eq3}
  I(\nu ,t, \vec{r})=\frac{\sqrt{3}\cdot e^2}{m\cdot
  c^3}\int_{0}^{E_{max}}dE\int dz\cdot N(E,t, \vec{r})\cdot
  H_{\bot}\cdot\frac{\nu}{\nu_s}\cdot\int_{\nu/\nu_s}^{\infty}K_{5/3}(\eta)d\eta,
\end{equation}

$$\nu_s=\frac{3}{4\pi}\cdot\frac{e H}{m c}\cdot\left(\frac{E}{m
c^2}\right)^2,$$ where  $H_\perp$ is the magnetic field projection
on the image plane, $\int dz$ is an integral along the line of
sight and $K_{5/3}$ is modified Bessel function. The numerical
calculations were made with help of the Mathematica 5.1 package.

\section*{Frequency dependence of synchrotron radiation intensity distribution}
\indent \indent The profiles of  synchrotron radiation intensity
distribution at different frequencies are given in Fig.2. Though the
remnant is not fully symmetrical we choose a sphere with $R=50pc$ as
approach. We neglect the motion of the shock front and consider the
SNR as an immovable injection source.

In Fig.2a $ \tau=3.6\cdot10^{11}s, l_{dif}=1.3\cdot10^{20}cm,
V_{dif}=3.7\cdot10^8cm/s$: the diffusion length is higher than SNR
radius, it almost completely defines the size of the radio image. In
Fig.2b $ \tau=8\cdot10^{10}s, l_{dif}=6.3\cdot10^{19}cm,
V_{dif}=7.8\cdot10^8 cm/s$, the life time and the diffusion length
decrease, and a gap (valley) appears in the figure. In Fig.2c the
life time and the diffusion length further decrease
$\tau=5\cdot10^{10}s, l_{dif}=5\cdot10^{19}cm,
V_{dif}=9\cdot10^8cm/s$, and the gap increases. In Fig.2d with
parameters $ \tau=4.2\cdot10^{10}s, l_{dif}=4.6\cdot10^{19}cm,
V_{dif}=1\cdot10^9 cm/s$ the gap reaches maximum value: the
diffusion length becomes less than own size of the system and has no
influence in its radio image. The approximation of a constant
diffusion factor is valid if the following equality is carried out:
${l_{dif1}}/{l_{dif2}}\approx\sqrt[4]{{\nu_2}/{\nu_1}}$. For the
chosen frequencies and corresponding  values of diffusion lengths
($l_{dif1}=1.3\cdot10^{20}cm, \nu_1=20 MHz$ and
$l_{dif2}=4.6\cdot10^{19}cm, \nu_2= 1420 MHz$) the equality is
carried out with a good accuracy
$1.3\cdot10^{20}cm/4.6\cdot10^{19}cm\approx\sqrt[4]{1420 MHz/20 MHz}
\approx 2.9$ It is necessary to note, that velocity of the SNR shock
front $R(t)=3\cdot10^7 cm/s$ can be of the same order as the
diffusion velocity and even more, therefore in general it is
necessary to consider the front movement.

\section*{Conclusions}
\indent \indent In the framework of the diffusion model we have
explained the observational data obtained by the UTR-2 according to
which the size of the supernova remnants on the low frequencies is
is higher than the one at high frequencies. It has been shown that
in our case the diffusion factor may be considered as a constant.

\section*{Acknowledgements}
The authors thank M.A. Sidorchuk for the image of SNR {HB3}.

\newpage

\textbf{Figure 1.} Brightness temperature contour map of SNR HB3 at
frequency 20 MHz by M.A. Sidorchuk and E.A. Abramenkov
\cite{1}.\vspace{10ex}

\textbf{Figure 2.} Profiles of intensity of synchrotron radiation
distribution for spherical source with constant diffusion factor for
following parameters a) $\nu=20 MHz;$ b) $\nu=408 MHz;$ c) $\nu =
1000 MHz;$ d) $\nu= 1420 MHz$, $I_\nu$ is expressed in relative
units in the same scale.\vspace{10ex}

Figures are available on YSC home page
(http://ysc.kiev.ua/abs/proc14$\_$8.pdf).


\begin{thebibliography}{5}
{\small
\bibitem{1}Abramenkov, E. A., Sidorchuk, M. A. Radiophysics and Radioastronomy, V. 11, 2, pp. 134-154 (2006)
\bibitem{2}Tian, W. W., Leahy, D. A. Astronomy and Astrophysics, V. 436, pp. 187-193 (2005)
\bibitem{3}Bannikova, E.Yu., Kontorovich, V.M., Radiophysics and Radioastronomy, V. 9, 1, pp. 29-36  (2004)
\bibitem{4}Bannikova, E.Yu., The journal of Kharkiv National University,
physical series. Nuclei, Particles, Fields, V. 627, pp. 57-62 (2005)
\bibitem{5}Ginzburg, V. L., "Theoretical Physics and Astrophysics",
Nauka, Moscow, 488 p. (1987)}
\end{thebibliography}
\end{document}